\documentclass[conference,letterpaper]{IEEEtran}

\usepackage[utf8]{inputenc}
\usepackage[T1]{fontenc}
\usepackage{amsmath}
\usepackage{amssymb}
\usepackage{booktabs}
\usepackage{graphicx}
\usepackage{subcaption}
\usepackage{tabularx}
\usepackage{makecell}
\usepackage{textcomp}
\usepackage{ragged2e}
\usepackage{xspace}
\usepackage[table,dvipsnames]{xcolor}
\usepackage{enumitem}
\usepackage{pbalance}

\newcolumntype{L}{>{\RaggedRight\arraybackslash}X}
\definecolor{tableheadspec}{HTML}{446688}

\usepackage[backend=biber,style=numeric-comp,doi=true,url=true,eprint=true,sorting=none]{biblatex}
\DeclareSourcemap{
  \maps[datatype=bibtex]{
    \map{
      \step[fieldset=abstract, null]
    }
  }
}
\AtBeginBibliography{%
  \footnotesize
  \setlength{\bibitemsep}{0pt}
  \setlength{\itemsep}{1pt}
}

\PassOptionsToPackage{hyphens}{url}\usepackage{hyperref}
\hypersetup{
    colorlinks=true,
    linkcolor=ForestGreen,
    citecolor=ForestGreen,
    urlcolor=RoyalBlue,
    pdftitle={Rolling Round-Robin Rate},
    pdfauthor={Mahesh Madhav},
    pdfsubject={Benchmarks},
    pdfkeywords={Standardization, Benchmarks, CPU Performance, Workload Characterization},
    pdfnewwindow=true,
    pdfdisplaydoctitle=true,
    bookmarksopen=true
}
\usepackage{cleveref}

\ifLuaTeX
\protected\def\pdfmapline {\pdfextension mapline}
\fi
\usepackage{fix-cm}    
  
\makeatletter
\newcommand\HUGE{\@setfontsize\Huge{37}{34}}
\makeatother  
  
\DeclareFontFamily{T1}{federation}{}
\DeclareFontShape{T1}{federation}{m}{n}{ <-> virtfed}{}
  
\pdfmapline{+federation < federation.ttf <T1-WGL4.enc}
  
\newcommand\tngfont[1]{{\usefont{T1}{federation}{m}{n} #1 }}

\begin{document}

\title{\HUGE{\tngfont{Rolling\,Round\,-Robin\,Rate:}}\\
\LARGE{Standard Heterogeneous Throughput for SPEC CPU}}

\author{
  \IEEEauthorblockN{Mahesh Madhav}
  \IEEEauthorblockA{\emph{Ampere Computing}}
  \IEEEauthorblockA{\emph{Portland, OR}}

  \and
  \IEEEauthorblockN{Christoph M\"{u}llner}
  \IEEEauthorblockA{\emph{VRULL}}
  \IEEEauthorblockA{\emph{Vienna, Austria}}
  
  \and
  \IEEEauthorblockN{Jiangning Liu}
  \IEEEauthorblockA{\emph{Ampere Computing}}
  \IEEEauthorblockA{\emph{Santa Clara, CA}}
  
  \and
  \IEEEauthorblockN{Philipp Tomsich}
  \IEEEauthorblockA{\emph{VRULL}}
  \IEEEauthorblockA{\emph{Vienna, Austria}}
  
  \and
  \IEEEauthorblockN{Jeff Baxter}
  \IEEEauthorblockA{\emph{Ampere Computing}}
  \IEEEauthorblockA{\emph{Santa Clara, CA}}
}

\maketitle
\newcommand{\Red}[1]{{\color{red} #1}}
\newcommand{\ignore}[1]{}
\newcommand{\asm}[1]{\texttt{#1}}
\newcommand{\sys}[1]{\texttt{#1}}
\newcommand{\kw}[1]{\textit{#1}}
\newcommand{\kwb}[1]{\textbf{#1}}
\newcommand{\type}[1]{\textit{#1}}
\newcommand{\Response}[1]{{\color{blue} #1}}
\newcommand{\XXX}[1]{\Red{\textbf{XXX[}#1\textbf{]}}}
\newcommand{\tocite}[1]{\Red{CITE:\cite{#1}}}
\newcommand{\toref}[1]{\Red{REF:\ref{#1}}}
\newcommand{\parasub}[1]{\smallskip\noindent\textit{{#1:}\xspace}}
\newcommand{\mahesh}[1]{\textcolor{purple}{#1}}
\newcommand{\anyone}[1]{\textcolor{blue}{#1}}
\newcommand{\niparagraph}[1]{\noindent\textbf{\textsf{#1}\hspace{0.5em}}}
\newcommand\TODO[1]{\textcolor{red}{TODO: #1}}

\newenvironment{CompactItemize}%
  {\begin{list}{$\blacktriangleright$}%
    {\leftmargin=\parindent \itemsep=2pt \topsep=2pt
     \parsep=0pt \partopsep=0pt}}%
  {\end{list}}
\renewcommand{\labelitemi}{$\blacktriangleright$}

\newcommand{\malloc}{{\texttt{Malloc}}}
\newcommand{\linklist}{{\textsf{Linked-List}}}

\begin{abstract}
SPEC CPU has long provided a common foundation for comparing processor,
compiler, memory-system, and platform performance. Its multi-copy SPECrate
mode measures homogeneous throughput by running many copies of the same
benchmark at once. That mode remains valuable, but modern cloud and server
systems commonly run heterogeneous collections of jobs whose interactions are
shaped by shared caches, memory bandwidth, power management, operating-system
scheduling, and noisy neighbors. SPEC CPU 2026 introduces Rolling Round-Robin
Rate (RRR), an exhibition run style that uses the existing rate suites to
generate deterministic heterogeneous multiprogrammed workloads. This paper
describes RRR as a benchmark methodology and proposes a SPEC-like scoring
model for future RRR reporting: compute per-benchmark average throughput and
coefficient of variation from per-copy ratios, compute a SPEC-style geometric
mean for each copy across the suite, and average that population of copy
geomeans to form a suite score with its own coefficient of variation. RRR
therefore preserves the familiar SPEC throughput tradition while exposing
richer information about variability, interference, and heterogeneous-system
behavior.
\end{abstract}

\section{Introduction}

The purpose of a benchmark is not only to run a program quickly. A useful
benchmark must define the work, control the measurement, validate the result,
and make comparisons reproducible. SPEC CPU has served this role for more than
three decades by providing portable source benchmarks, well-defined run rules,
reference workloads, validation, and standardized metrics for speed and
throughput \cite{cpu2000,cpu2006,cpu2017,john_suite_growth}. Its results are
used by CPU buyers, system vendors, compiler writers, and academic researchers
because the benchmark harness turns complex software into a common measurement
experiment.

SPECrate measures throughput by running multiple copies of one benchmark at a
time. This homogeneous capacity method dates to the early SPEC CPU suites
\cite{spec_homogenous_capacity}. It remains an important stress case; if a
machine has a shared-cache bottleneck, memory-bandwidth limit, thermal limit,
or scheduler issue, many synchronized copies of the same benchmark can expose
it. Homogeneous rate also has a clean interpretation. Each benchmark produces a
normalized throughput score, and the suite score is the geometric mean of those
scores, avoiding the well-known problems of arithmetic means for normalized
benchmark ratios \cite{dont_lie}.

However, homogeneous rate is no longer the only throughput question that
matters. Modern high-core-count servers and cloud systems are routinely used by
many tenants, services, build jobs, virtual machines, and data-processing tasks
at the same time. Their behavior depends not just on single-program speed or
single-benchmark saturation, but on interference among unlike programs. A cache
partitioning policy, SMT pairing policy, power limit, frequency governor, or OS
scheduler may look reasonable under homogeneous load and behave differently
when compute-bound, memory-bound, branch-heavy, and front-end-heavy programs
run together. Researchers have repeatedly constructed their own multiprogrammed
workloads to study these effects, but the field lacks a widely adopted,
standard workload construction method for full-application heterogeneous CPU
rate measurements.

SPEC CPU 2026 introduces Rolling Round-Robin Rate (RRR), a new exhibition run
style for this purpose \cite{spec_cpu_2026_next_generation,rrrrate, rrr_article}. RRR runs
the SPEC rate benchmarks in a deterministic rolling schedule: every copy runs
the same selected roster, but each copy starts at a different point and then
walks through the roster. The result is a controlled heterogeneous workload in
which every benchmark receives equal exposure across copies, while the
concurrent mix changes over time.

This paper makes three contributions. First, it motivates RRR as a standard
methodology for heterogeneous multiprogrammed SPEC CPU runs. Second, it defines
a SPEC-like RRR scoring model that could be added to official RRR reporting:
compute a mean throughput ratio and coefficient of variation for each
benchmark, compute a geometric mean for each copy across the suite, and average
that population of copy geomeans for the suite score. Third, it argues that
RRR can produce a richer result set than homogeneous rate because the same
measurements yield coefficient-of-variation diagnostics at both the benchmark
and suite levels. These diagnostics are directly useful for noisy-neighbor
studies, OS scheduling, heterogeneous-core evaluation, resource partitioning,
and reproducibility analysis.

\section{Why Heterogeneous Rate Needs a Standard}

\begin{figure*}[!th]
    \centering 

    \begin{tabular}{@{}c@{\hfill}c@{\hfill}c@{}}

        \begin{subfigure}[b]{0.313\textwidth}
            \includegraphics[width=\linewidth]{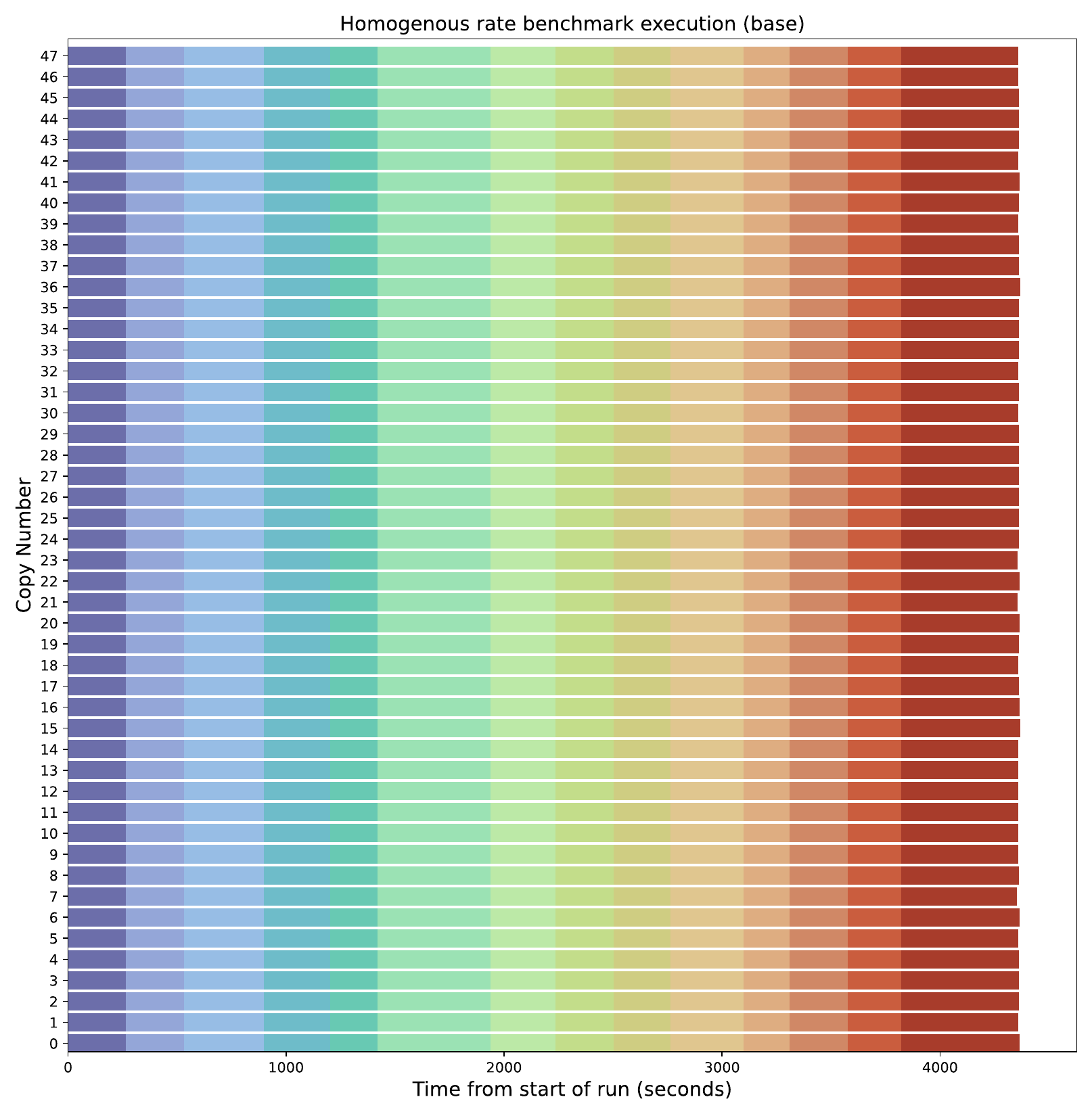}
            \caption{Standard refrate}
            \label{fig:sub_a}
        \end{subfigure}
        & 
        
        \begin{subfigure}[b]{0.313\textwidth}
            \includegraphics[width=\linewidth]{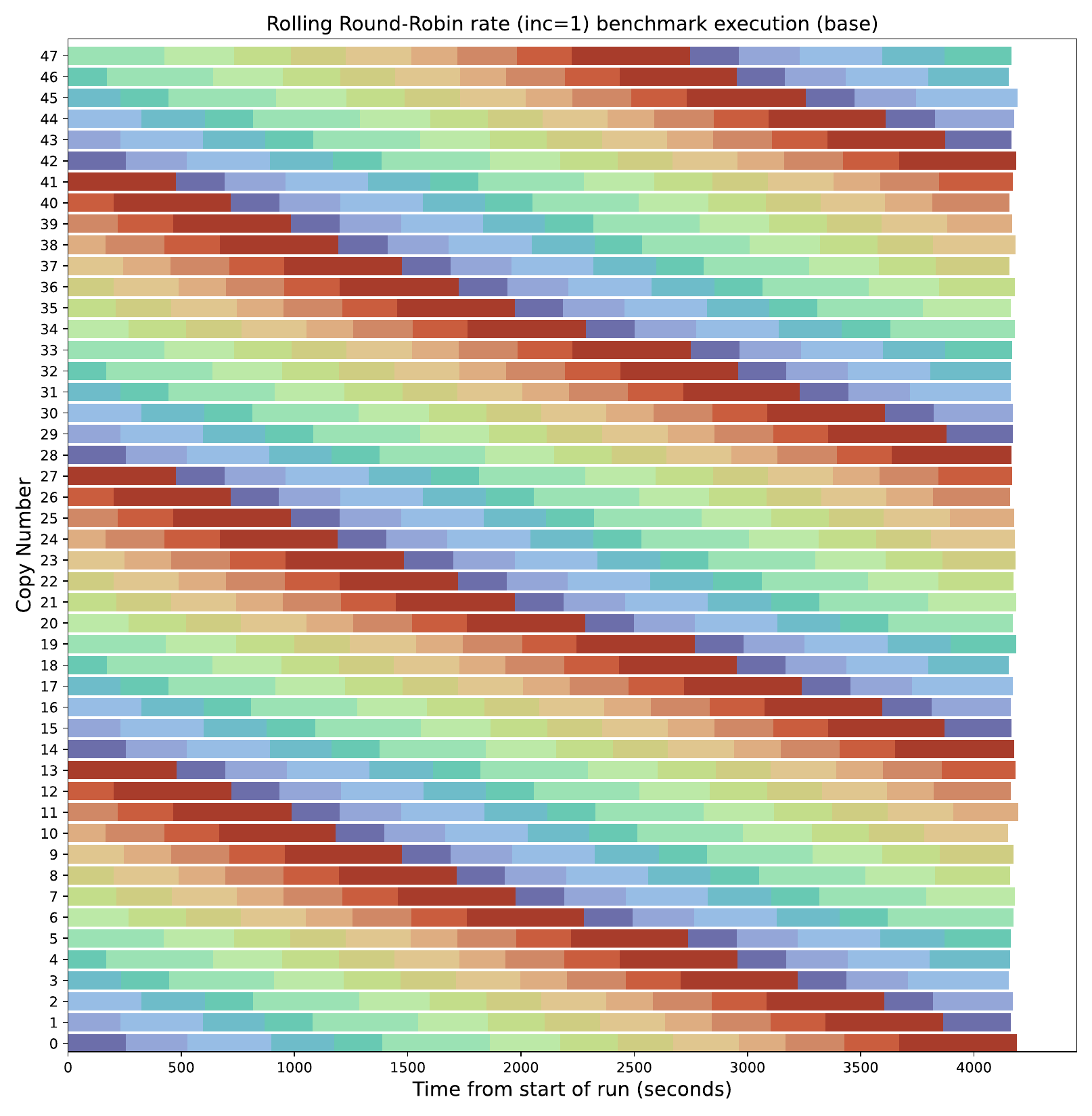}
            \caption{rolling round-robin rate, inc=1}
            \label{fig:sub_b}
        \end{subfigure}
        & 
        
        \begin{subfigure}[b]{0.3775\textwidth}
            \includegraphics[width=\linewidth]{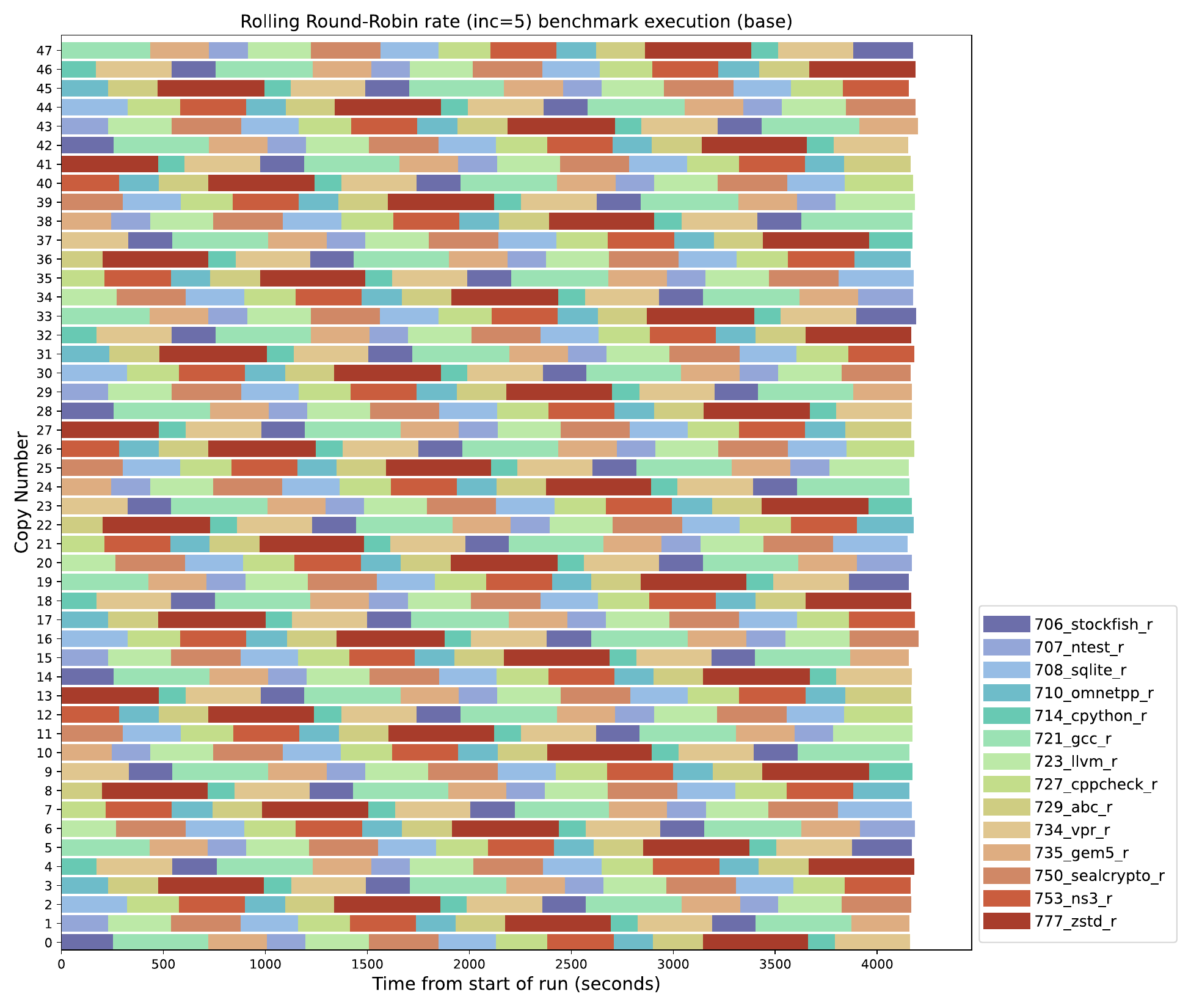}
            \caption{rolling round-robin rate, inc=5}
            \label{fig:sub_c}
        \end{subfigure}
        
    \end{tabular} 

    \caption{Time plots showing all 14 integer rate benchmarks executing on schedules for homogeneous (refrate) and heterogeneous (rrrrate) on 48 copies. Every copy runs the same
    selected work; homogeneous rate aligns copies of one benchmark, while RRR
    rolls copies through the roster to create deterministic heterogeneous load. The ``inc'' parameter tells the scheduler how to increment through the roster of benchmarks, allowing for some customization in the heterogeneous scheduling. Plots were generated with the rate\_timeplot.py script found in the scripts.misc directory bundled with the CPU2026 distribution.}
    \label{fig:rrrplots}
\end{figure*}

The literature on multiprogrammed workload evaluation shows both a sustained
need for heterogeneous workload experiments and a lack of convergence on how to
construct them. A single-program benchmark is relatively easy to define -- it is
a program, an input, a build, and a measurement rule. A multiprogrammed
benchmark is more complicated because each program has its own runtime and
phase behavior, and the relative alignment of those phases changes contention
for shared resources. Jacobvitz et al. formalized this problem with a
four-tuple definition and showed that different multiprogram benchmark
definitions built from the same programs can produce drastically different
conclusions \cite{hetero_bench4}. FIESTA made a related distinction between
sample imbalance, which comes from unequal standalone runtimes, and schedule
imbalance, which comes from real asymmetric contention during concurrent
execution \cite{multiprogram_workload}. A good heterogeneous rate method should
control sample imbalance without erasing the schedule imbalance that it intends
to measure.

Computer-architecture studies have addressed the construction problem in many
ways. Some work samples benchmark combinations, stratifies mixes, or uses
principal component and cluster analysis to choose representative
multiprogrammed workloads for simulation \cite{hetero_bench2,hetero_bench3}.
Other work identifies representative program-input pairs or observes that many
benchmark suites overrepresent similar bottlenecks
\cite{designing_workloads,picking_inputs,only_four_bottlenecks,benchmark_characterization_1991}.
SPEC CPU itself is heavily studied for suite balance, subsetting \cite{lizy_subsetting,lizy_subsetting2,lizy_2017,cpu2017_characterization}, and behaviors of applications, memory, and energy \cite{cpu2017_energy_characterization,cpu2017_memory_centric,cpu2017_another_mem_hier_charz,cpu2006_17_memsystem}.
These valuable studies also show that workload selection is a
methodological choice with consequences.

The need becomes even clearer in work that studies resource sharing directly.
Researchers have used custom built benchmark mixes to evaluate cache partitioning, SMT pairing,
resource scheduling, colocation fairness, prefetcher selection, and
noisy-neighbor isolation
\cite{phase_aware_cache_partitioning,SMT_pairs,dynamic_adaptive_power_multicore,contention_aware_job_colocation, micro_mama, resource_partitioning_in_colocation,sdcbench}.
OS and runtime systems work similarly needs concurrent workloads to evaluate
thread allocation, asymmetric multicore scheduling, and malleable application
management
\cite{efficient_multicore_proc_scheduling,hetero_aware_os_scheduler_for_asymmetric_multicore,workload_splitting_heterocore}.
Recent work also predicts concurrent execution time or cache behavior using
performance counters \cite{predicting_concurrent_execution_time,predicting_cache_behavior_concurrent}.
Across these areas, the research question is often not whether heterogeneous
interference exists, but how to create a workload that others can reproduce and
compare against.

Several proposals attempt to make multiprogrammed SPEC-style evaluation faster
or more statistically tractable. SPECcast and BenchCast annotate regions of
interest and synchronize representative slices so that full-program behavior can
be approximated at lower cost \cite{hetero_speccast,hetero_bench1}. Those
methods are useful when simulation cost dominates, but they deliberately change
the execution to a selected region. RRR takes the complementary approach for
real-system measurement, by running complete SPEC rate benchmarks through the
standard harness and using a deterministic schedule to create a heterogeneous
mix. This preserves full-application interactions, including the cross-effects
that may be inconvenient but are exactly what shared-resource systems must
handle.

Metric choice is a separate issue. Multiprogrammed systems can be evaluated
using system throughput, weighted speedup, harmonic means, normalized turnaround
time, fairness indices, or objective-specific metrics
\cite{hetero_bench5,hetero_bench6,survey_of_multiprogram,four_metrics_to_evaluate_multicore}.
RRR does not require the community to settle every fairness question before it
can be useful. It provides a standard experiment and a natural throughput score,
while also exposing distributions that support fairness, variability, and
isolation studies.

\section{Rolling Round-Robin Rate}

RRR is a heterogeneous run style for SPEC CPU rate suites. Given a selected
benchmark roster $B$ with $N$ benchmarks and a copy count $M$, each copy runs
all selected benchmarks sequentially. The copies use the same deterministic
roster but begin at different positions. If copy 0 starts with benchmark
$b_0$, copy 1 starts with $b_1$, and so on cyclically through the roster. After
each benchmark finishes, the copy advances through the roster until every copy
has run every selected benchmark once. All verification is delayed until the end of all $N\times M$ benchmarks have completed; the harness does not insert any barriers or synchronization points while the roster is executing.

The CPU 2026 harness exposes this mode through \texttt{--rrrrate}; the
\texttt{--rrrrate\_inc} option controls how each copy advances through the
roster \cite{rrrrate}. For example, with six benchmarks $(a,b,c,d,e,f)$ and
\texttt{inc=2}, copy 0 runs $(a,c,e,b,d,f)$ while copy 1 runs
$(b,d,f,a,c,e)$. Positive increments create different deterministic
heterogeneous schedules. The special case \texttt{inc=0} keeps each copy on its
initial benchmark, which is useful for quick validation but is not the
heterogeneous schedule of interest.

RRR has four useful properties. First, it is deterministic. A published RRR
experiment can specify the suite, copy count, selected benchmarks, and increment
value, allowing the schedule to be reconstructed. Second, it gives equal
exposure. Every benchmark runs once on every copy, so benchmark representation
does not depend on a random mix or an arbitrary pairing. Third, it focuses on
CPU behavior. SPEC CPU benchmarks are selected and hardened to reduce unrelated
sources of nondeterminism such as excessive file I/O, unstable random behavior,
and benchmark self-measurement
\cite{cpu2006_fileio,cpu2006_cxx,cpu2006_training,cpu2006_perfmon,cpu2006_hotspots}.
Fourth, it preserves full-program interactions. Unlike kernel extraction or
ROI-only schemes, the benchmark instances run through the standard validated
workloads, so phase changes and long-range behavior remain part of the
experiment.

\section{A SPEC-like RRR Score}

RRR can support a primary score that stays close to the SPEC tradition while
adding richer diagnostics. Although RRR is an exhibition mode and the
current formatted reports emphasize average time and time CV, the underlying
RRR measurements also support a natural throughput score. We propose extending
RRR reporting with four ratio-based quantities: \texttt{ratio\_avg},
\texttt{ratio\_cv}, \texttt{basemean\_avg}, and \texttt{basemean\_cv}. These
quantities are proposed additions to the current SPEC CPU 2026 exhibition
output. An example result is shown in \autoref{tab:rrr_ratio_scores}.

Let $R_b$ be the reference time for benchmark $b$. Let $T_{b,c,i}$ be the
elapsed time for benchmark $b$ when it runs on copy $c$ during RRR iteration
$i$, for $M$ copies and $I$ iterations. The normalized rate ratio for one
execution is
\begin{equation}
  x_{b,c,i} = M \cdot \frac{R_b}{T_{b,c,i}} .
\end{equation}
For each benchmark and copy, define $x_{b,c}$ as the copy's representative
ratio. With one iteration this is simply $x_{b,c,1}$; with multiple iterations
it follows the usual SPEC practice of selecting the median of three runs for
that copy. The benchmark-level RRR throughput score is the arithmetic mean of
those per-copy ratios:
\begin{equation}
  \mathrm{ratio\_avg}_b =
    \bar{x}_b =
    \frac{1}{M}\sum_{c=1}^{M} x_{b,c}.
\end{equation}
The physical interpretation of ratio\_avg is the aggregate delivered throughput in reference-runs per unit time. Then, its coefficient of variation is computed over the same population of copy
ratios:
\begin{equation}
  \mathrm{ratio\_cv}_b =
    \frac{\sqrt{\frac{1}{M}\sum_{c=1}^{M}
    (x_{b,c}-\mathrm{ratio\_avg}_b)^2}}{\mathrm{ratio\_avg}_b}.
\end{equation}
The current average time and time CV outputs remain useful companion
diagnostics, but ratio-based quantities are the natural throughput metrics for
scoring.

  \begin{table}[t]
  \centering
  \caption{ratio\_avg scores and CV for refrate and RRR in two configs, corresponding to the runs depicted in \autoref{fig:rrrplots}. CV is reported as a percentage. RRR's rolling roster replaces self-contention with a mixed crowd, so a bandwidth-bound benchmark's RRR ratio should exceed its homogeneous-rate ratio, while compute-bound benchmarks should be nearly indifferent between the two modes. The per-benchmark RRR-to-rate ratio shift is therefore an in-band memory-boundedness index, obtained from one RRR run plus the standard rate result. This standardises a classification that otherwise requires per-benchmark performance-counter triage. ratio\_cv adds a complementary mix-sensitivity signal.}
  \label{tab:rrr_ratio_scores}
  
  \setlength{\tabcolsep}{2.2pt}
  \begin{tabular}{l@{\hspace{1em}}rr@{\hspace{2em}}rr@{\hspace{2em}}rr}
  \toprule
  Benchmark & \multicolumn{2}{c}{refrate} & \multicolumn{2}{c}{rrrrate-inc=1} & \multicolumn{2}{c}{rrrrate-inc=5} \\
  \cmidrule(lr){2-3}\cmidrule(lr){4-5}\cmidrule(lr){6-7}
   & ratio & CV\% & ratio & CV\% & ratio & CV\% \\
  \midrule
  706.stockfish & 154.0 & 0.46 & 158.7 & 0.45 & 158.3 & 0.79 \\
  707.ntest     & 116.6 & 0.29 & 113.4 & 0.36 & 113.4 & 0.22 \\
  708.sqlite    &  82.7 & 0.36 &  82.2 & 0.47 &  82.1 & 0.58 \\
  710.omnetpp   &  96.8 & 1.37 & 101.0 & 0.93 & 101.1 & 0.72 \\
  714.cpython   & 124.2 & 0.39 & 123.7 & 0.55 & 123.6 & 0.41 \\
  721.gcc       &  80.4 & 0.46 &  84.3 & 0.91 &  84.3 & 0.87 \\
  723.llvm      & 103.5 & 0.35 &  98.1 & 0.59 &  98.1 & 0.73 \\
  727.cppcheck  &  77.0 & 0.69 &  79.7 & 0.85 &  79.8 & 0.83 \\
  729.abc       &  94.7 & 0.45 &  98.8 & 0.55 &  98.8 & 0.64 \\
  734.vpr       &  88.8 & 0.86 &  96.0 & 0.77 &  95.9 & 0.83 \\
  735.gem5      & 158.9 & 0.46 & 158.4 & 0.26 & 158.3 & 0.33 \\
  750.sealcrypt & 103.7 & 0.13 & 104.3 & 0.29 & 104.3 & 0.23 \\
  753.ns3       & 193.8 & 0.48 & 190.6 & 0.44 & 190.7 & 0.35 \\
  777.zstd      &  79.3 & 0.53 &  82.4 & 1.00 &  82.5 & 0.80 \\
  \midrule
  Geomean       & 106.6 & --   & 108.2 & 0.29 & 108.1 & 0.28 \\
  \bottomrule
  \end{tabular}
  \end{table}

The suite score should preserve complete copy-level observations before
averaging them away. For each copy, compute a SPEC-style geometric mean across
the benchmark ratios observed by that copy:
\begin{equation}
  G_c =
    \left(\prod_{b \in B} x_{b,c}\right)^{1/|B|}.
\end{equation}
The proposed headline suite score, SPECrrrrate, is the arithmetic
mean of the per-copy geomeans:
\begin{equation}
  \mathrm{basemean\_avg} =
    S_{\mathrm{RRR}} =
    \frac{1}{M}\sum_{c=1}^{M} G_c .
\end{equation}
This order of operations stays close to SPEC practice: each $G_c$ is a
geometric mean of normalized rate ratios across the suite
\cite{dont_lie,cpu2017}. Averaging the $G_c$ values then treats the copies as a
population of equal-exposure suite observations. To be clear, the suite score is NOT the geomean of the benchmark scores; rather it is the average of the geomeans of each copy.

The same population provides the suite-level coefficient of variation:
\begin{equation}
  \mathrm{basemean\_cv} =
    \frac{\sqrt{\frac{1}{M}\sum_{c=1}^{M}
    (G_c-\mathrm{basemean\_avg})^2}}{\mathrm{basemean\_avg}}.
\end{equation}
With this extension, RRR would yield useful statistics at two levels. At the
benchmark level, \texttt{ratio\_avg} and \texttt{ratio\_cv} would show mean
throughput and relative spread for each benchmark under heterogeneous co-runner
contexts. At the suite level, \texttt{basemean\_avg} and
\texttt{basemean\_cv} would show the mean and relative spread of complete
copy-level suite geomeans. These four quantities allow a result to answer both
``how much throughput did the system deliver?'' and ``how evenly and
predictably did it deliver that throughput under heterogeneous load?'' On
nominally homogeneous systems, the CVs can expose placement effects, frequency
effects, or noisy-neighbor sensitivity. On heterogeneous systems, they may
reflect core type, thermal policy, OS scheduling, or resource partitioning.

\section{Limitations}

RRR has limitations that should be stated explicitly. First, the rolling schedule does not keep the
  machine at a perfectly constant heterogeneous load for the entire run. As shorter benchmark instances
  finish, later-running instances can become "late-comers" that execute under lighter system load. The
  round-robin construction spreads this effect across the roster, but the spreading is exact only when the
  benchmark count and copy count align cleanly. 
  
  RRR's mixing depends on the increment. For a roster of N benchmarks and
  increment k, the modular walk visits every benchmark before repeating only when
  gcd(k,N)=1. Relatively prime increments therefore give the most diverse roster
  rotation, while increments that share a factor with N repeat alignments sooner
  and reduce co-runner diversity.

Finally, RRR's equal-exposure property is count-based, not time-weighted. Each selected benchmark runs
  once on every copy, but longer-running benchmarks naturally overlap with more phases and peer benchmarks
  than shorter-running ones. This preserves full-application behavior, but means RRR should be interpreted
  as a standardized heterogeneous throughput experiment, not a perfectly balanced pairwise co-runner matrix.

\section{Uses for Industry and Academia}

RRR is useful because it standardizes a workload shape that many communities
already need. For industry, it provides a repeatable mixed-load throughput
experiment for CPU buyers, server vendors, cloud providers, operating system teams, compiler
teams, and silicon designers. Homogeneous rate remains the right tool for
single-benchmark saturation, but RRR adds a controlled way to ask how a system
behaves when different applications contend for shared CPU resources. The same RRR
run can support average-throughput comparison, per-benchmark sensitivity
analysis, and variability analysis for noisy-neighbor behavior.

For academia, RRR offers a common baseline for studies that currently build
custom workload mixes. Operating system process schedulers can oversubscribe the system and compare policies using the same
benchmark roster and increment values. Heterogeneous-core research can report
whether work is distributed effectively across core types. Resource
partitioning and colocation studies can evaluate both throughput and
dispersion. Compiler and architecture studies can examine whether an
optimization improves the average but increases relative spread, or whether it
improves stability without changing the headline score. Because the run uses
full SPEC applications, results remain connected to a widely studied benchmark
suite whose behavior has been characterized over multiple generations
\cite{specint_analysis,intel_cpu2000_06,alberta_workloads}.

RRR has already been used this way in early SPEC CPU 2026 characterization
work. Li et al. use RRR both as a full-suite heterogeneous reference and as a
composition primitive for targeted proxy workloads \cite{cpu2026_characterization_rrr}.
By selecting complementary SPEC CPU 2026 benchmarks and running them in RRR
mode, they construct reproducible mixes that bracket selected microarchitectural pressure points to mimic the response of DCPerf \cite{dcperf}. Their study also makes an
important distinction, that RRR is not a full quality-of-service benchmark for
datacenter co-location, but it is a practical way to generate controlled
heterogeneous CPU pressure from real benchmark programs. This is exactly the
kind of flexible, reproducible use case that a standard heterogeneous rate mode
is intended to enable.

RRR should therefore be viewed as an addition to the benchmarking toolkit, not
a replacement for existing modes. SPECspeed answers a latency-oriented
single-workload question. SPECrate answers a homogeneous throughput question.
SPECrrrrate answers a heterogeneous throughput question and exposes the
variability around that answer. These headline scores are not interchangeable
and should not be compared directly since each stresses the processor differently
and answers a different performance question. Taken together, these modes give a more
complete view of CPU performance in the modern many-core era.

\section{Conclusion}

Rolling Round-Robin Rate extends SPEC CPU throughput measurement to a
standardized heterogeneous setting. Its schedule is intentionally simple. Every copy runs
every selected benchmark, but the copies roll through the roster from different
starting points. That simplicity is its strength. It controls workload
construction, preserves full-application execution, and allows for a measurement
that can be conducted similarly across systems and studies.

The proposed RRR reporting extension also supports a practical score. By
averaging normalized throughput for each benchmark across copies, and by
computing a SPEC-style geomean for each copy across the suite, RRR stays close
to the SPEC scoring tradition of ratios against a reference machine. Averaging the population of copy geomeans gives
a practical suite score, while the coefficient of variance spread of those geomeans exposes information
that homogeneous rate largely hides:
benchmark sensitivity to co-runners, copy-level balance, and relative spread
around the final score. We expect these diagnostics to be useful for studying
noisy neighbors, OS scheduling, heterogeneous systems, resource partitioning,
and future benchmark methodology. RRR gives the community a common experiment
for mixed-workload CPU throughput and a richer basis for comparing modern
systems.

\printbibliography

\end{document}